\documentclass[reprint,
superscriptaddress,
 amsmath,amssymb,
 aps,pra,
]{revtex4-1}
\usepackage{multirow}
\usepackage[caption=false,justification=justified]{subfig}
\usepackage{leftidx}
\usepackage{graphicx}% Include figure files
\usepackage{dcolumn}% Align table columns on decimal point
\usepackage{bm}% bold math
\usepackage{hyperref}% add hypertext capabilities
\usepackage{rotating}
\usepackage{xcolor}

\newcommand{\pt}[1]{\left( #1 \right)}
\newcommand{\pq}[1]{\left[ #1 \right]}
\newcommand{\pg}[1]{\left\lbrace #1 \right\rbrace}
\newcommand{\bra}[1]{\left\langle #1 \right\vert}
\newcommand{\ket}[1]{\left\vert #1 \right\rangle}
\newcommand{\braket}[2]{\left\langle #1 \vert #2 \right\rangle}

\begin{document}
\preprint{APS/123-QED}

\title{Multi-parameter estimation of\\ the state of two interfering photonic qubits}% Force line breaks with \\

\author{Luca Maggio}
\affiliation{School of Mathematics and Physics, University of Portsmouth}
\author{Danilo Triggiani}
\affiliation{School of Mathematics and Physics, University of Portsmouth}
\author{Paolo Facchi}
\affiliation{Dipartimento di Fisica, Universit\`{a} di Bari, I-70126 Bari, Italy}
\affiliation{INFN, Sezione di Bari, I-70126 Bari, Italy}
\author{Vincenzo Tamma}
\email{vincenzo.tamma@port.ac.uk}
\affiliation{School of Mathematics and Physics, University of Portsmouth}
\affiliation{Institute of Cosmology and Gravitation, University of Portsmouth}

\date{\today}% It is always \today, today,

\begin{abstract}
It is demonstrated a two-photon interfering technique based on polarization-resolved measurements for the simultaneous estimation with the maximum sensitivity achievable in nature of multiple parameters associated with the polarization state of two interfering photonic qubits. This estimation is done by exploiting a novel interferometry technique based on polarization-resolved two-photon interference. We show the experimental feasibility and accuracy of this technique even when a limited number of sampling measurements is employed. This work is relevant for the development of quantum technologies with photonic qubits  and sheds  light on the physics at the interface between multiphoton interference, boson sampling, multi-parameter quantum sensing and quantum information processing.
\end{abstract}

\maketitle

Several interferometry-based protocols have been recently introduced for high-precision estimation of a single parameter encoded in a two-photon state, e.g. for estimating optical length, time delay~\cite{hong1987measurement, lyons2018attosecond, chen2019hong, triggiani2023ultimate}, polarization~\cite{Harnchaiwat:20, photonics10010072, PhysRevApplied.19.014008}, and transverse displacement~\cite{triggiani2023estimation}. These new technologies are based on two-photon interferometry, that takes place when two identical photons impinge on two distinct input ports of a balanced beam splitter~\cite{hong1987measurement, shih1988new, bouchard2020two}. In this case, the two photons will bunch at the same output channel, due to the interference between the two two-photon amplitudes associated with the two two-photon paths leading to coincidence events. Instead, for non-identical photons, the probability of a bunching event depends on the overlap between the single-photon states. In principle, by encoding an unknown parameter in this overlap, it is possible to estimate it by counting the bunching rate. The efficiency of such a scheme can be evaluated by means of quantum metrology techniques~\cite{a61aa5fe-d74a-3133-bed2-f35c3c555015, holevo2011probabilistic, rohatgi2015introduction}, and it depends on the overlap between the two single-photon states: the larger is the overlap, the better is the precision~\cite{Harnchaiwat:20, knoll2019role, knoll2023simultaneous}. Also, the precision of the measurement can increase by exploiting sampling measurements which resolve photonic parametes which are conjugate to the ones one wish to estimate. This idea has been proven useful for the estimation of time delay and of transverse displacement~\cite{triggiani2023estimation,triggiani2023ultimate}. In both cases, the ultimate precision in the estimation has been reached.

An important problem in quantum sensing is the estimation of polarization. So far, the only technique that aims to estimate polarization of single photons based on two-photon interference hinges on measurements unable to retrieve the full photonic quantum metrological information. This scheme, which has been investingated in visible~\cite{Harnchaiwat:20} and telecom wavelength range~\cite{photonics10010072} has two main limitations. First, the protocol has been developed for the estimation of the polar angle of the polarization and do not resolve the relative phase. Second, it does not reach the ultimate precision even in estimating a single parameter. 

In this Letter we introduce a new technique which enables to estimate multiple polarization parameters, of interfering single photon states, particularly the polar angle and the relative phase, with the highest precision achievable in nature and without the need of resolving any conjugated parameter. This quantum sensing result has application in quantum communication and quantum computation, and it allows to achieve the ultimate precision for the estimation of the relative phase of two photonic qubits for any  polar angle.

\begin{figure}
\centering
  \includegraphics[width=80mm]{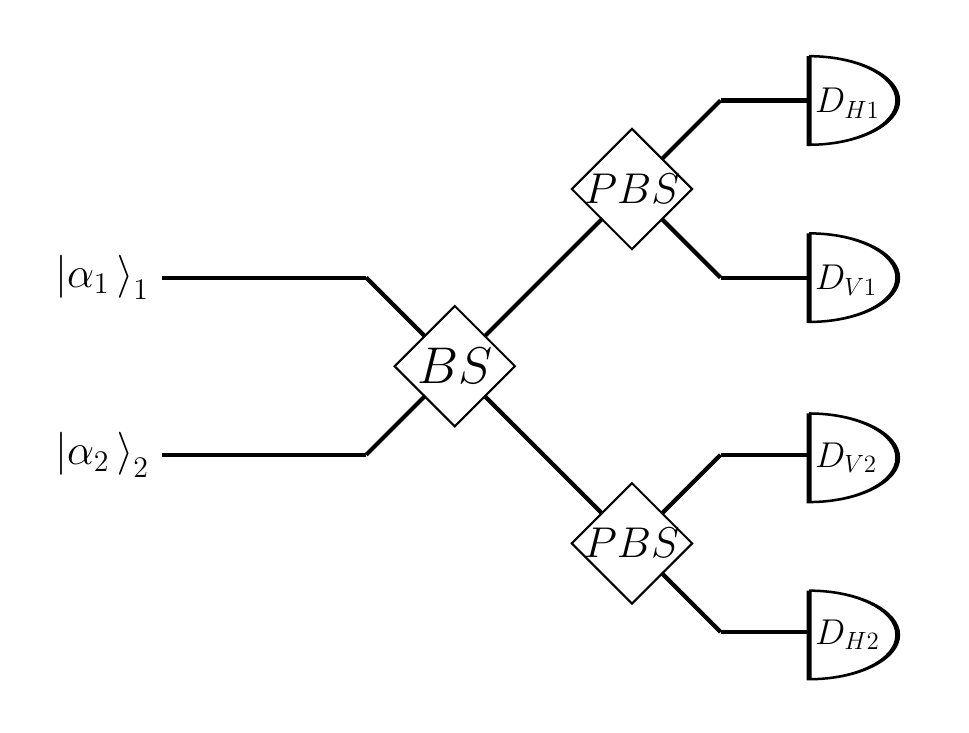}
  \caption{Representation of the sensing scheme. Two photons with two polarization states $\ket{\alpha_1}_1$ and $\ket{\alpha_2}_2$ shown in~\eqref{eq:InitState} enter in the input channel of a 50:50 beam splitter (BS). Then, their polarization is resolved with two polarizing beam splitters (PBSs) and four different detectors.}
  \label{fig:SetUp2024}
\end{figure}

\paragraph*{The scheme} 
The sensing scheme proposed in this Letter is represented in \figurename~\ref{fig:SetUp2024}. Here, $\ket{\alpha_i}_i$, with $i=1,2$, is the state of the photon of polarization $\alpha_i$ injected in the $i$-th input port of a balanced beam splitter (BS), with
\begin{equation}
\begin{split}
\ket{\alpha_1}_1&=\cos{\frac{\theta}{2}}\ket{H}_1+\sin{\frac{\theta}{2}}e^{i\phi_1}\ket{V}_1\, ,\\
\ket{\alpha_2}_2&=\cos{\frac{\theta}{2}}\ket{H}_2+\sin{\frac{\theta}{2}}e^{i\phi_2}\ket{V}_2.
\end{split}
\label{eq:InitState}
\end{equation}
Two polarizing beam splitters (PBS) are connected to the output channels of the BS, and they project the two single-photon states onto the base $\{\ket{H},\ket{V}\}$. Finally, four single photon detectors are connected to the outputs of the PBSs.

The possible outputs of this scheme can be arranged in four groups.  A \emph{coincidence} event occurs when each photon impinges on a different PBS. It will be labelled as $C$ and happens when one of the following couples of detectors clicks: $\pt{D_{H_1}\,,D_{H_2}}$, $\pt{D_{H_1}\,,D_{V_2}}$, $\pt{D_{V_1}\,,D_{H_2}}$, $\pt{D_{V_1}\,,D_{V_2}}$.  A \emph{single bunching} event occurs when both photons impinge on the same PBS and two different detectors click. It will be labelled as $SB$ and happens when one of the following couples of detectors clicks: $\pt{D_{H_1}\,,D_{V_1}}$, $\pt{D_{H_2}\,,D_{V_2}}$.  \emph{Double bunching}  events, $DB_{H}$ or $DB_{V}$,  occur if a single detector clicks and both photons are detected with \emph{horizontal} polarization ($DB_{H}$), or \emph{vertical} polarization ($DB_{V}$). In these two cases one of the horizontal detectors $D_{H_i}$, or of the vertical detectors $D_{V_i}$, clicks respectively, with $i=1,2$. These four outputs have, respectively, the probabilities (see Appendix~\ref{app:Probs})
\begin{align}
\begin{split}
P_{C}=\frac{1}{2}\sin^2\theta\sin^2\delta_\phi,\\
P_{SB}=\frac{1}{2}\sin^2\theta\cos^2\delta_\phi,\\
P_{DB_{H}}=\pt{\frac{1+\cos\theta}{2}}^2,\\
P_{DB_{V}}=\pt{\frac{1-\cos\theta}{2}}^2,\\
\end{split}\label{eq:Probs}
\end{align}
where
\begin{equation}
\delta_\phi=\left\vert\frac{\phi_1-\phi_2}{2}\right\vert. 
\end{equation}

The measurement protocol is based on sampling the four different outcomes from these probabilities. For each sample, $N$ is the number of repetitions of the experiment and $N_{X}$, $X=C,SB,DB_{H},DB_{V}$ are the numbers of event registered for each type.
The unknown parameters $\theta$ and $\delta_\phi$ are thus estimated through the maximum-likelihood estimators as (see Appendix~\ref{app:Likelihood})
\begin{align}
\begin{split}
\tilde{\theta}&=\arccos \pt{\frac{N_{DB_{H}}-N_{DB_{V}}}{N}},\\
\tilde{\delta}_\phi&=\arctan\sqrt{\frac{N_C}{N_{SB}}},
\end{split}
\end{align}
with $\tilde{\theta}\in\pq{0,\pi}$ and $\tilde{\delta}_\phi\in\pq{0,\pi/2}$.

\paragraph*{Bounds on the precision}

Since this is a two-parameter estimation problem, a $2\times 2$ covariance matrix $\mathrm{Cov}[{\tilde{\theta},\tilde{\delta}_\phi}]$ controls the precision in the estimation of $\theta$ and $\delta_\phi$, and, in particular, its diagonal terms $\sigma^2_{\tilde{\theta}}$ and $\sigma^2_{\tilde{\delta}_\phi}$ represent the variances of the estimators $\tilde{\theta}$ and $\tilde{\delta}_\phi$, respectively. To analyse the precision in the estimation of these two parameters, the following bounds in the covariance matrix are used:
\begin{equation}
   \mathrm{Cov}\pq{\tilde{\theta},\tilde{\delta}_\phi}\geq \frac{F^{-1}\pt{\theta,\delta_\phi}}{N}\geq\frac{H^{-1}\pt{\theta, \delta_\phi}}{N}.\label{eq:QCRB}
\end{equation}
Here, $F\pt{\theta,\delta_\phi}$ is the Fisher information matrix and $H\pt{\theta, \delta_\phi}$ is the quantum Fisher information matrix. The first inequality is the Cram\'{e}r-Rao bound (CRB), which holds for unbiased estimators. It defines the maximum information achievable from the scheme presented, and it can be saturated in the asymptotic regime of large $N$~\cite{a61aa5fe-d74a-3133-bed2-f35c3c555015}. The second inequality instead is the quantum Cram\'{e}r-Rao bound (QCRB), that defines the maximum precision achievable by \emph{any} possible measurement scheme. In other words, if $F\pt{\theta,\delta_\phi}=H\pt{\theta, \delta_\phi}$, the scheme is maximally efficient for an asymptotically large number $N$.

\begin{figure}
\centering
  \includegraphics[width=80mm]{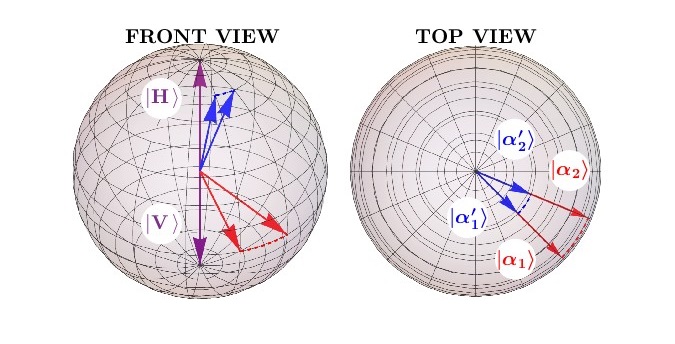}
  \caption{The distance between two points on the Bloch sphere with different phases and same polar angle is bigger when the two points are near the equator rather than near the poles. In figure, the points near the equator are associated with $\ket{\alpha_1}$ and $\ket{\alpha_2}$. The points near the poles are associated with $\ket{\alpha'_1}$ and $\ket{\alpha'_2}$. The greater is the distance between the states, the easier is to distinguish them. The possibility to distinguish two states with the same polar angle is related to the sensitivity on the difference in the  phases.}
  \label{fig:Bloch}
\end{figure}

The matrices $H\pt{\theta, \delta_\phi}$ for the probe state and $F\pt{\theta,\delta_\phi}$ for this measurement protocol are evaluated in Appendices~\ref{app:A} and~\ref{app:Fisher} respectively, where it is shown that
\begin{equation}
F\pt{\theta, \delta_\phi}=H\pt{\theta, \delta_\phi}=2\begin{pmatrix}
1&0\\
0&\sin^2\theta
\end{pmatrix},\label{QFIMi}
\end{equation}
namely, that the proposed scheme saturates the QCRB. 

Interestingly, since this matrix is diagonal, it is possible to retrieve $\theta$ and $\delta_\phi$ independently. Also, the precision in the estimation of $\delta_\phi$ is linked to the value of the polar angle of the states $\ket{\alpha_1}_1$ and $\ket{\alpha_2}_2$: this is an effect related to the geometry of the Bloch sphere. In fact, a small variation of the relative phase (either $\phi_1$ or $\phi_2$) provides a greater change in the state if its Bloch vector is at the equator, rather than at the poles. This means that the maximum sensitivity to the value of the relative phase can be obtained for value of $\theta$ close to $\pi/2$~\cite{brivio2010experimental}, as it is shown in \figurename~\ref{fig:Bloch}.

To the best of the authors' knowledge, the saturation of the QCRB for the estimation of $\delta_\phi$ for arbitrary values of $\theta$ has never been proven so far for any scheme. In addition, the scheme here proposed allows to simultaneously estimate also $\theta$ with the maximum precision. 
It is worth to notice that, in this optimal measurement scheme, the polarisation states $\ket{\alpha_i}_i$ in Eq.~\eqref{eq:InitState} of the two non-identical photons are completely indistinguishable at the detectors, in the sense that the two photons yield the same probabilities to be observed in the base $\{\ket{H},\ket{V}\}$ of the PBS, namely
$|{_1}\!\braket{\zeta}{\alpha_1}_1|^2=|{_2}\!\braket{\zeta}{\alpha_2}_2|^2$ with $\zeta=H,V$.
Indeed, achieving indistinguishability between the photons at the detectors by choosing the appropriate basis for the measurement is an optimal choice for two-photon interference sensing techniques~\cite{triggiani2023ultimate,triggiani2023estimation}

\paragraph*{Accuracy and precision in the non-asymptotic regime}
\begin{figure}
\centering
        \includegraphics[width=80mm]{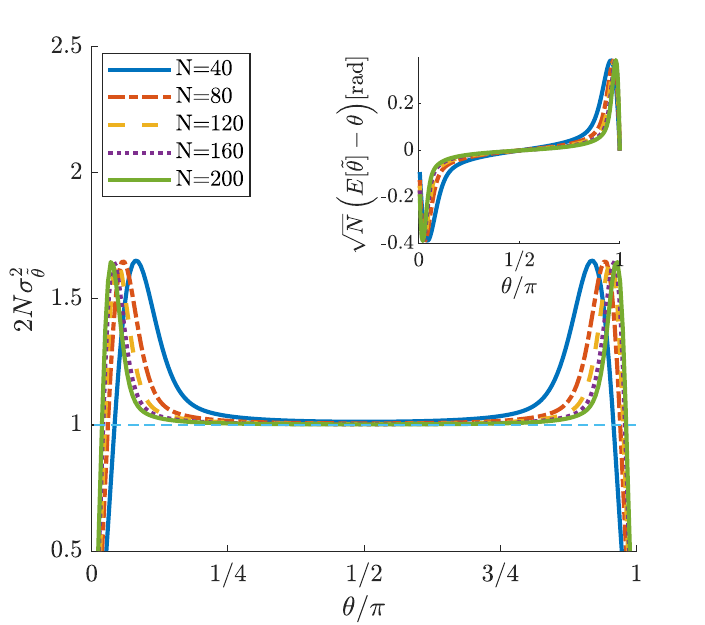}
        \caption{Saturation of the CRB (see Eqs.~\ref{eq:QCRB} and~\ref{QFIMi}) for the maximum-likelihood estimator $\tilde{\theta}$ as a function of $\theta$ by increasing the number $N$ of experimental runs. In the inset, it is shown the bias in the estimation as a function of $\theta$ by increasing $N$.  One can see that $\tilde{\theta}$ is unbiased for $\theta=0,\pi/2,\pi$ independently of the value $N$. For values close to $\theta= 0,\pi$ the estimator is biased and closer to the poles than the actual value $\theta$, but the bias can be reduced by acquiring more data (notice that in the plots the peaks of the bias appear to scale as $1/\sqrt{N}$). There is no dependence on the value of $\delta_\phi$.}
        \label{fig:CRB}
\end{figure}
\begin{figure}
\centering
        \includegraphics[width=80mm]{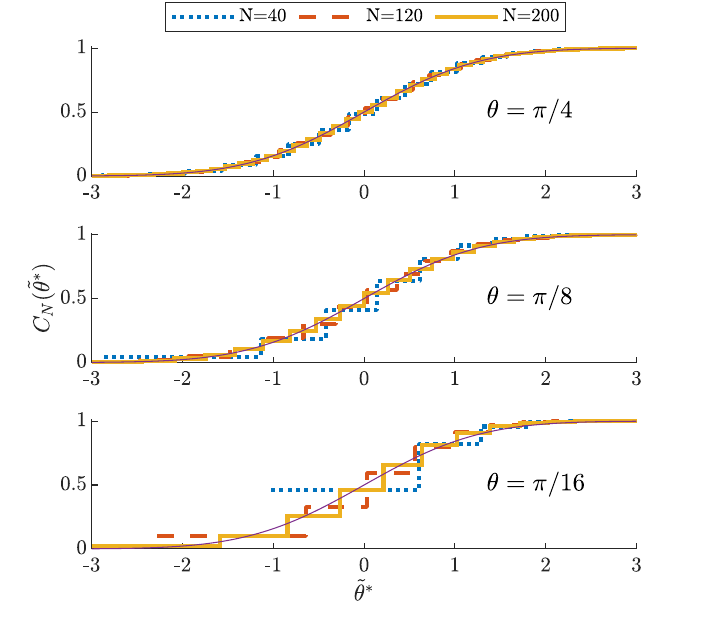}
        \caption{Cumulative distribution of $\tilde{\theta}^*=({\theta-E[{\tilde{\theta}}]})/{\sigma_{\tilde{\theta}}}$ for different values of $\theta$ and $N$ and the cumulative distribution of the normal distribution (solid line), where $E[\tilde{\theta}]=\sum_{\tilde{\theta}}\tilde{\theta}P(\tilde{\theta}\vert N)$. The probability $P(\tilde{\theta}\vert N)$ is binomial. $P(\tilde{\theta}\vert N)$ and its cumulative are evaluated in Appendix~\ref{app:Prob}. Fixing $N$, the interval of the distributions shrinks even below $3\sigma_{\tilde{\theta}}$ the more $\theta$ tends to $0$. The Gaussianity can be restored by increasing $N$.}
        \label{fig:CRBCum}

\end{figure}
\begin{figure}
\centering
        \includegraphics[width=80mm]{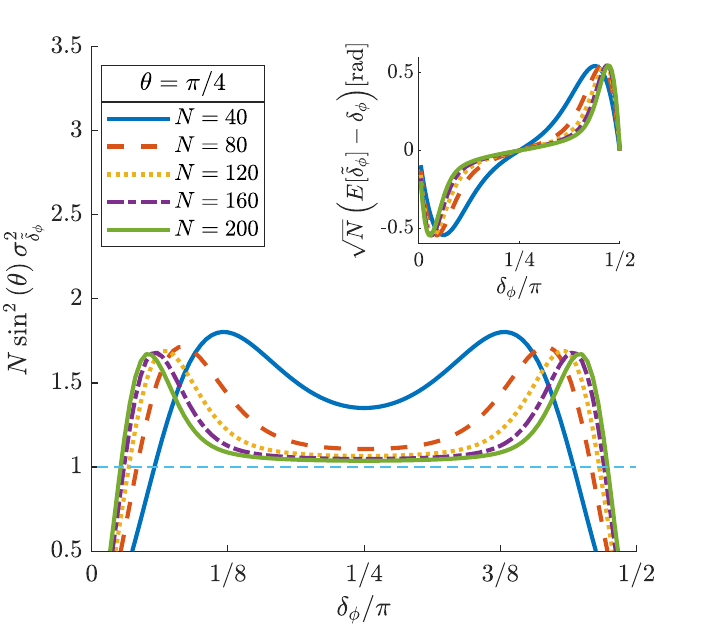}
        \includegraphics[width=80mm]{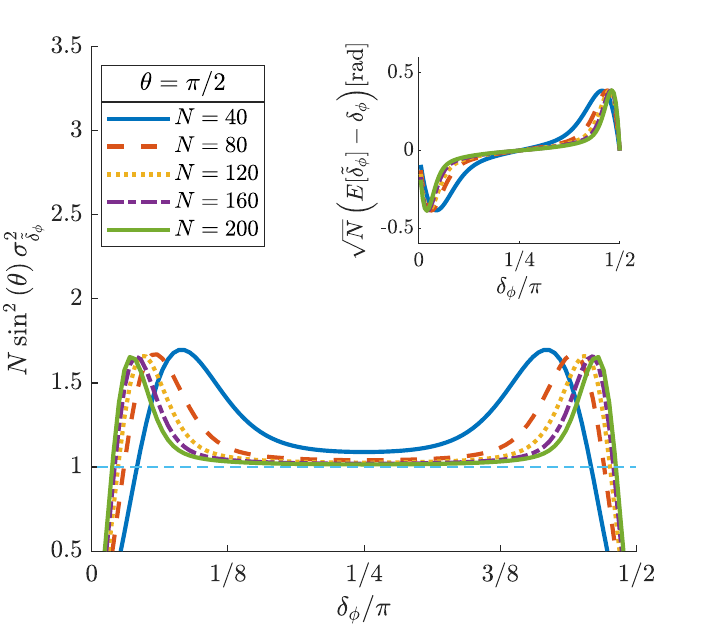}
        \caption{Saturation of the CRB (see Eqs.~\ref{eq:QCRB} and~\ref{QFIMi}) for the maximum-likelihood estimator $\tilde{\delta}_\phi$ as a function of $\delta_\phi$ for different values of $N$. In the inset, it is shown the bias in the estimation as a function of $\delta_\phi$ for different values of $N$. One can see that $\tilde{\delta}_{\phi}$ is unbiased for $\delta_{\phi}=0,\pi/4,\pi/2$ independently of the value $N$. For values close to $\theta= 0,\pi/2$ the estimator is biased, but the bias can be reduced by acquiring more data (notice that in the plots the peaks of the bias appear to scale as $1/\sqrt{N}$). The plots show that the more the Bloch vectors of the probe state are near the equator, the better is the estimation (both in precision and accuracy) of $\delta_\phi$. This happens because close to the poles the estimation fails, as it is discussed in the text and in Appendix~\ref{app:A}.}
        \label{fig:CRBde}
\end{figure}
\begin{figure}
\centering
        \includegraphics[width=80mm]{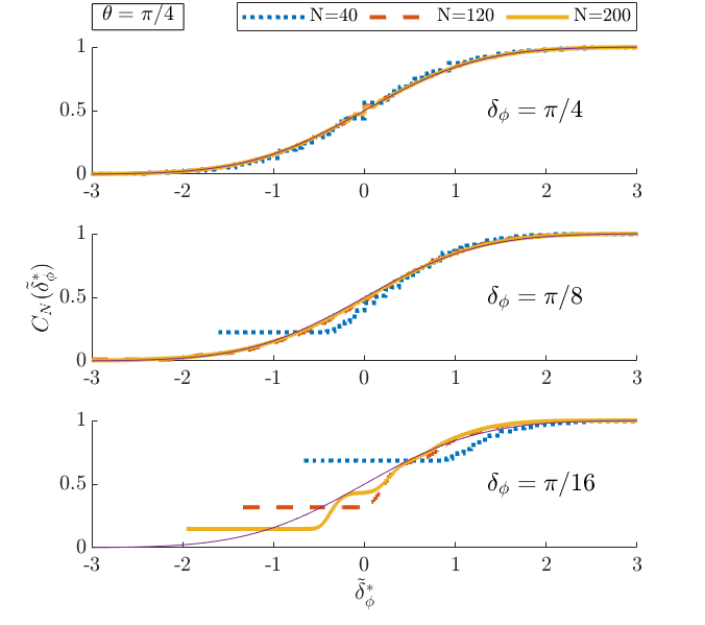}
        \includegraphics[width=80mm]{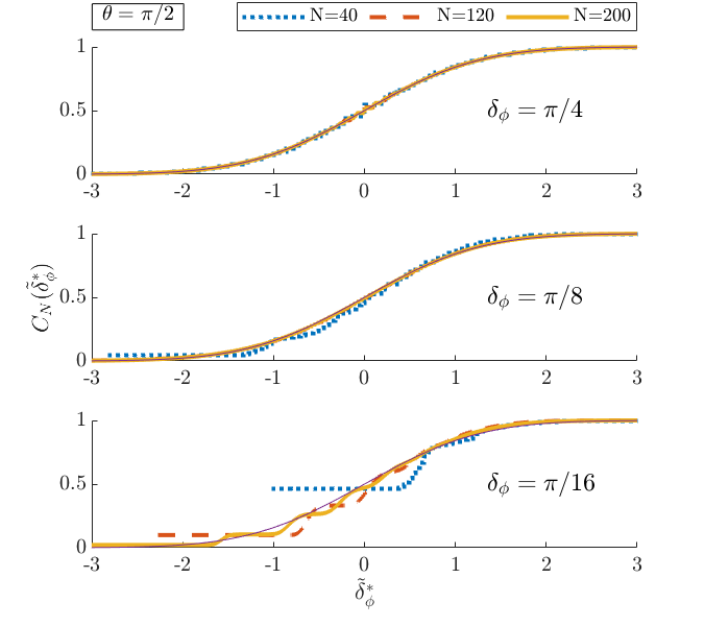}
        \caption{Cumulative distribution of the estimator $\tilde{\delta}_\phi$ by changing the parameters $\theta$, $\delta_\phi$ and $N$. The cumulative distribution is plotted as a function of $\tilde{\delta}_\phi^*=(\tilde{\delta}_\phi-E[{\tilde{\delta}_\phi}])/\sigma_{\tilde{\delta}_\phi}$, where $E[\tilde{\delta}_\phi]=\sum_{\tilde{\delta}_\phi}\tilde{\delta}_\phi P(\tilde{\delta}_\phi\vert N)$. The probability $P(\tilde{\delta}_\phi \vert N)$ and its cumulative are discussed in Appendix~\ref{app:Prob}. The cumulative distributions are compared with the normal standard cumulative distribution (the continuous line). Fixing $N$, the interval of the distributions shrinks even below $3\sigma_{\tilde{\delta}_{\phi}}$ the more $\delta_{\phi}$ tends to $0$. The Gaussianity can be restored by increasing $N$}
        \label{fig:CRBCumde}

\end{figure}

Although the CRB is saturated in the asymptotic regime of large $N$, a finite number $N$ of sampling measurements from the probability distribution in Eq.~\eqref{eq:Probs} suffices to reach accuracy and optimal precision. Indeed, similarly to inner-mode resolved boson sampling~\cite{tamma2023scattershot,laibacher2015physics,tamma2016multi,tamma2015sampling,10.1145/1993636.1993682,tamma2021boson}, although here in the context of quantum metrology, the output probability distribution does not need to be experimentally reproduced avoiding therefore the need of a large number of experimental runs. This can be shown numerically by evaluating the bias and the variance normalized to the CRB for both the estimators $\tilde{\theta}$ and $\tilde{\delta}_\phi$ by using the probability distributions of the likelihood estimators as a function of N obtained in Appendix~\ref{app:Prob}.

 In the main figure of \figurename~\ref{fig:CRB}, it is shown the variance of $\tilde{\theta}$ normalized to the CRB (see Eqs.~\ref{eq:QCRB} and~\ref{QFIMi}), namely $2N\sigma^2_{\tilde{\theta}}$, vs the parameter $\theta$ for different values of $N$. One can see that for $\theta=0,\pi$, the variance $\sigma^2_{\tilde{\theta}}$ tends to zero, as the output of the protocol is deterministic. In fact, for $\theta=0,\pi$ coincidence events or single bunching events cannot happen since the photons are identical. Therefore, only the events $DB_{H}$ for $\theta=0$ and $DB_{V}$ for $\theta=\pi$ can happens, respectively with probability $P_{DB_{H/V}}=1$. Also, the larger the value of $N$, the larger is the interval centred at $\theta=\pi/2$ for which the CRB can be considered saturated. Moving outside this interval, the precision decreases until it reaches a minimum point (maximum value of $2N\sigma^2_{\tilde{\theta}}$). To better understand this behaviour we recall that in the asymptotic regime of large $N$, the one for which the CRB is saturated, the distribution of $\tilde{\theta}$ must be approximately Gaussian. It is possible to show that the distribution of $\tilde{\theta}$ is not in the regime of gaussianity in the interval of $\theta$ for which the CRB is not saturated (see Appendix~\ref{app:Prob} for an analysis of the probability distributions of $\tilde{\theta}$ and $\tilde{\delta}_\phi$). Nonetheless, it is possible to approach the CRB saturation for most value of the parameters $\theta$ and $\delta_\phi$ and with a number $N$ of experimental iterations only of the order of 100.
 
 In \figurename~\ref{fig:CRBCum}, the cumulative function of the normal distribution is compared with the cumulative function of the variable $\tilde{\theta}^*=({\theta-E[{\tilde{\theta}}]})/{\sigma_{\tilde{\theta}}}$, which has zero mean and unit variance, for any value of $N$. One can see that, for a fixed $N$, the more $\theta$ approaches the boundaries, the less the cumulative distributions of $\tilde{\theta}^*$ resemble the one of a Gaussian. However, increasing the value of $N$, and thus reducing $\sigma_{\tilde{\theta}}$, restores the Gaussian shape of the distribution of $\tilde{\theta}^*$.

In \figurename~\ref{fig:CRBde}, it is shown the variance of $\tilde{\delta}_{\phi}$ normalized to its CRB (see Eqs.~\ref{eq:QCRB} and~\ref{QFIMi}), namely $2N\sin^2\theta\sigma^2_{\tilde{\delta}_\phi}$, vs the parameter $\delta_\phi$ for different values of $N$. It is worth noticing that $2N\sin^2\theta\sigma^2_{\tilde{\delta}_\phi}$ depends on both $\delta_\phi$ and $\theta$. In particular, the variance will tend faster to the CRB the closer $\theta$ is to $\pi/2$. In fact, as shown in \figurename~\ref{fig:CRBCumde}, the cumulative distribution resembles sooner a Gaussian cumulative distribution when $\theta=\pi/2$. However, both the saturation of the CRB and the bias for $\tilde{\delta}_\phi$ have similar properties as the ones shown in \figurename~\ref{fig:CRB}. As before, when the parameter is at the edge of its interval of definition ($\delta_\phi=\pq{0,\pi/2}$), the variance is zero. The CRB is reached more efficiently in the middle of the interval $\tilde{\delta}_\phi=\pi/4$, where the estimation is also unbiased. This condition can always be reached independently of the real parameter $\delta_\phi$ by inserting in the second output channel of the balanced beam splitter a phase shift gate described by the operator $\hat{U}_{SG}$, so that $\ket{\alpha_2}_2$ is transformed in
 \begin{align}
   \ket{\alpha_{2,\epsilon}}_2=\hat{U}_{SG}\ket{\alpha_2}_2
   &=\cos{\frac{\theta}{2}}\ket{H}_2+\sin{\frac{\theta}{2}}e^{i\pt{\phi_2+\epsilon}}\ket{V}_2\,,
 \end{align}
 where $\epsilon$ is fixed. By choosing a proper $\epsilon$ for which $\delta_\phi-\epsilon/2\approx\pi/4$ it is possible to increase the performance of the protocol. In this case the estimator of the parameter $\delta_{\phi}$ will assume the form:
 \begin{equation}
     \tilde{\delta}_{\phi}=\arctan\sqrt{\frac{N_C}{N_{SB}}}+\frac{\epsilon}{2}.
 \end{equation}

\paragraph*{Conclusion}
We presented a feasible and scalable polarization-resolved interferometry technique for the estimation of multiple parameters encoded in a two-photon polarization state. This scheme allows to estimate two parameters, both the polar angle $\theta$ and the  relative phase $\delta_\phi$ by reaching the ultimate quantum precision. This result has been proved by means of quantum metrology tecnhiques, i.e. by evaluating the Fisher information of the scheme and the quantum Fisher information of the input state. The saturation of the CRB in the proposed technique is intimately connected with the two-photon indistinguishability at the detectors obtained through polarization-resolved sampling measurements, without the need of solving any conjugated parameter. Also, we investigated the precision and the accuracy of the measurement in the non-asymptotic regime. Remarkably, already for a number $N$ of sampling measurements of the order of 100 our scheme is accurate and saturate the CRB bound for most of the values of $\theta$ and $\delta_\phi$. Furthermore, we have shown  that it is possible to optimize the precision of $\delta_\phi$ by adapting the polarization state in input to a suitable range of values to be estimated. To the best of the authors' knowledge, the estimation of $\delta_\phi$ by reaching the ultimate precision has never been proven so far with any other sensing scheme. Even if beyond the scope of this current work, this research can also inspire further investigations of the proposed sensing scheme in the case of non efficient detectors and photonic states which may differ in parameters other than their polarization. In conclusion, these remarkable results at the interface between quantum metrology, quantum optical interference and boson sampling, can pave the way to new high-precision sensing techniques. They also lay the foundation for future  schemes based on multi-parameter interferometry techniques for the estimation of polarization parameters that involve more than two qubits in more general quantum information networks.
\acknowledgments
This project is partially supported by Xairos Systems Inc. VT also acknowledges partial support from the Air Force Office of Scientific Research under award number FA8655-23-17046. PF was partially supported by Istituto Nazionale di Fisica Nucleare (INFN) through the project ``QUANTUM'', by the Italian National Group of Mathematical Physics (GNFM-INdAM), and by the Italian funding within the ``Budget MUR - Dipartimenti di Eccellenza 2023--2027'' - Quantum Sensing and Modelling for One-Health (QuaSiModO).

\nocite{*}
\bibliography{Article1}

\begin{thebibliography}{26}%
\makeatletter
\providecommand \@ifxundefined [1]{%
 \@ifx{#1\undefined}
}%
\providecommand \@ifnum [1]{%
 \ifnum #1\expandafter \@firstoftwo
 \else \expandafter \@secondoftwo
 \fi
}%
\providecommand \@ifx [1]{%
 \ifx #1\expandafter \@firstoftwo
 \else \expandafter \@secondoftwo
 \fi
}%
\providecommand \natexlab [1]{#1}%
\providecommand \enquote  [1]{``#1''}%
\providecommand \bibnamefont  [1]{#1}%
\providecommand \bibfnamefont [1]{#1}%
\providecommand \citenamefont [1]{#1}%
\providecommand \href@noop [0]{\@secondoftwo}%
\providecommand \href [0]{\begingroup \@sanitize@url \@href}%
\providecommand \@href[1]{\@@startlink{#1}\@@href}%
\providecommand \@@href[1]{\endgroup#1\@@endlink}%
\providecommand \@sanitize@url [0]{\catcode `\\12\catcode `\$12\catcode `\&12\catcode `\#12\catcode `\^12\catcode `\_12\catcode `\%12\relax}%
\providecommand \@@startlink[1]{}%
\providecommand \@@endlink[0]{}%
\providecommand \url  [0]{\begingroup\@sanitize@url \@url }%
\providecommand \@url [1]{\endgroup\@href {#1}{\urlprefix }}%
\providecommand \urlprefix  [0]{URL }%
\providecommand \Eprint [0]{\href }%
\providecommand \doibase [0]{http://dx.doi.org/}%
\providecommand \selectlanguage [0]{\@gobble}%
\providecommand \bibinfo  [0]{\@secondoftwo}%
\providecommand \bibfield  [0]{\@secondoftwo}%
\providecommand \translation [1]{[#1]}%
\providecommand \BibitemOpen [0]{}%
\providecommand \bibitemStop [0]{}%
\providecommand \bibitemNoStop [0]{.\EOS\space}%
\providecommand \EOS [0]{\spacefactor3000\relax}%
\providecommand \BibitemShut  [1]{\csname bibitem#1\endcsname}%
\let\auto@bib@innerbib\@empty
%</preamble>
\bibitem [{\citenamefont {Hong}\ \emph {et~al.}(1987)\citenamefont {Hong}, \citenamefont {Ou},\ and\ \citenamefont {Mandel}}]{hong1987measurement}%
  \BibitemOpen
  \bibfield  {author} {\bibinfo {author} {\bibfnamefont {C.-K.}\ \bibnamefont {Hong}}, \bibinfo {author} {\bibfnamefont {Z.-Y.}\ \bibnamefont {Ou}}, \ and\ \bibinfo {author} {\bibfnamefont {L.}~\bibnamefont {Mandel}},\ }\href@noop {} {\bibfield  {journal} {\bibinfo  {journal} {Physical review letters}\ }\textbf {\bibinfo {volume} {59}},\ \bibinfo {pages} {2044} (\bibinfo {year} {1987})}\BibitemShut {NoStop}%
\bibitem [{\citenamefont {Lyons}\ \emph {et~al.}(2018)\citenamefont {Lyons}, \citenamefont {Knee}, \citenamefont {Bolduc}, \citenamefont {Roger}, \citenamefont {Leach}, \citenamefont {Gauger},\ and\ \citenamefont {Faccio}}]{lyons2018attosecond}%
  \BibitemOpen
  \bibfield  {author} {\bibinfo {author} {\bibfnamefont {A.}~\bibnamefont {Lyons}}, \bibinfo {author} {\bibfnamefont {G.~C.}\ \bibnamefont {Knee}}, \bibinfo {author} {\bibfnamefont {E.}~\bibnamefont {Bolduc}}, \bibinfo {author} {\bibfnamefont {T.}~\bibnamefont {Roger}}, \bibinfo {author} {\bibfnamefont {J.}~\bibnamefont {Leach}}, \bibinfo {author} {\bibfnamefont {E.~M.}\ \bibnamefont {Gauger}}, \ and\ \bibinfo {author} {\bibfnamefont {D.}~\bibnamefont {Faccio}},\ }\href@noop {} {\bibfield  {journal} {\bibinfo  {journal} {Science advances}\ }\textbf {\bibinfo {volume} {4}},\ \bibinfo {pages} {eaap9416} (\bibinfo {year} {2018})}\BibitemShut {NoStop}%
\bibitem [{\citenamefont {Chen}\ \emph {et~al.}(2019)\citenamefont {Chen}, \citenamefont {Fink}, \citenamefont {Steinlechner}, \citenamefont {Torres},\ and\ \citenamefont {Ursin}}]{chen2019hong}%
  \BibitemOpen
  \bibfield  {author} {\bibinfo {author} {\bibfnamefont {Y.}~\bibnamefont {Chen}}, \bibinfo {author} {\bibfnamefont {M.}~\bibnamefont {Fink}}, \bibinfo {author} {\bibfnamefont {F.}~\bibnamefont {Steinlechner}}, \bibinfo {author} {\bibfnamefont {J.~P.}\ \bibnamefont {Torres}}, \ and\ \bibinfo {author} {\bibfnamefont {R.}~\bibnamefont {Ursin}},\ }\href@noop {} {\bibfield  {journal} {\bibinfo  {journal} {npj Quantum Information}\ }\textbf {\bibinfo {volume} {5}},\ \bibinfo {pages} {43} (\bibinfo {year} {2019})}\BibitemShut {NoStop}%
\bibitem [{\citenamefont {Triggiani}\ \emph {et~al.}(2023)\citenamefont {Triggiani}, \citenamefont {Psaroudis},\ and\ \citenamefont {Tamma}}]{triggiani2023ultimate}%
  \BibitemOpen
  \bibfield  {author} {\bibinfo {author} {\bibfnamefont {D.}~\bibnamefont {Triggiani}}, \bibinfo {author} {\bibfnamefont {G.}~\bibnamefont {Psaroudis}}, \ and\ \bibinfo {author} {\bibfnamefont {V.}~\bibnamefont {Tamma}},\ }\href@noop {} {\bibfield  {journal} {\bibinfo  {journal} {Physical Review Applied}\ }\textbf {\bibinfo {volume} {19}},\ \bibinfo {pages} {044068} (\bibinfo {year} {2023})}\BibitemShut {NoStop}%
\bibitem [{\citenamefont {Harnchaiwat}\ \emph {et~al.}(2020)\citenamefont {Harnchaiwat}, \citenamefont {Zhu}, \citenamefont {Westerberg}, \citenamefont {Gauger},\ and\ \citenamefont {Leach}}]{Harnchaiwat:20}%
  \BibitemOpen
  \bibfield  {author} {\bibinfo {author} {\bibfnamefont {N.}~\bibnamefont {Harnchaiwat}}, \bibinfo {author} {\bibfnamefont {F.}~\bibnamefont {Zhu}}, \bibinfo {author} {\bibfnamefont {N.}~\bibnamefont {Westerberg}}, \bibinfo {author} {\bibfnamefont {E.}~\bibnamefont {Gauger}}, \ and\ \bibinfo {author} {\bibfnamefont {J.}~\bibnamefont {Leach}},\ }\href {\doibase 10.1364/OE.382622} {\bibfield  {journal} {\bibinfo  {journal} {Opt. Express}\ }\textbf {\bibinfo {volume} {28}},\ \bibinfo {pages} {2210} (\bibinfo {year} {2020})}\BibitemShut {NoStop}%
\bibitem [{\citenamefont {Sgobba}\ \emph {et~al.}(2023)\citenamefont {Sgobba}, \citenamefont {Pallotti}, \citenamefont {Elefante}, \citenamefont {Dello~Russo}, \citenamefont {Dequal}, \citenamefont {Siciliani~de Cumis},\ and\ \citenamefont {Santamaria~Amato}}]{photonics10010072}%
  \BibitemOpen
  \bibfield  {author} {\bibinfo {author} {\bibfnamefont {F.}~\bibnamefont {Sgobba}}, \bibinfo {author} {\bibfnamefont {D.~K.}\ \bibnamefont {Pallotti}}, \bibinfo {author} {\bibfnamefont {A.}~\bibnamefont {Elefante}}, \bibinfo {author} {\bibfnamefont {S.}~\bibnamefont {Dello~Russo}}, \bibinfo {author} {\bibfnamefont {D.}~\bibnamefont {Dequal}}, \bibinfo {author} {\bibfnamefont {M.}~\bibnamefont {Siciliani~de Cumis}}, \ and\ \bibinfo {author} {\bibfnamefont {L.}~\bibnamefont {Santamaria~Amato}},\ }\href {\doibase 10.3390/photonics10010072} {\bibfield  {journal} {\bibinfo  {journal} {Photonics}\ }\textbf {\bibinfo {volume} {10}} (\bibinfo {year} {2023}),\ 10.3390/photonics10010072}\BibitemShut {NoStop}%
\bibitem [{\citenamefont {Tsujimoto}\ \emph {et~al.}(2023)\citenamefont {Tsujimoto}, \citenamefont {Ikuta}, \citenamefont {Wakui}, \citenamefont {Kobayashi},\ and\ \citenamefont {Fujiwara}}]{PhysRevApplied.19.014008}%
  \BibitemOpen
  \bibfield  {author} {\bibinfo {author} {\bibfnamefont {Y.}~\bibnamefont {Tsujimoto}}, \bibinfo {author} {\bibfnamefont {R.}~\bibnamefont {Ikuta}}, \bibinfo {author} {\bibfnamefont {K.}~\bibnamefont {Wakui}}, \bibinfo {author} {\bibfnamefont {T.}~\bibnamefont {Kobayashi}}, \ and\ \bibinfo {author} {\bibfnamefont {M.}~\bibnamefont {Fujiwara}},\ }\href {\doibase 10.1103/PhysRevApplied.19.014008} {\bibfield  {journal} {\bibinfo  {journal} {Phys. Rev. Appl.}\ }\textbf {\bibinfo {volume} {19}},\ \bibinfo {pages} {014008} (\bibinfo {year} {2023})}\BibitemShut {NoStop}%
\bibitem [{\citenamefont {Triggiani}\ and\ \citenamefont {Tamma}(2024)}]{triggiani2023estimation}%
  \BibitemOpen
  \bibfield  {author} {\bibinfo {author} {\bibfnamefont {D.}~\bibnamefont {Triggiani}}\ and\ \bibinfo {author} {\bibfnamefont {V.}~\bibnamefont {Tamma}},\ }\href@noop {} {\bibfield  {journal} {\bibinfo  {journal} {Physical Review Letters}\ }\textbf {\bibinfo {volume} {132}},\ \bibinfo {pages} {180802} (\bibinfo {year} {2024})}\BibitemShut {NoStop}%
\bibitem [{\citenamefont {Shih}\ and\ \citenamefont {Alley}(1988)}]{shih1988new}%
  \BibitemOpen
  \bibfield  {author} {\bibinfo {author} {\bibfnamefont {Y.}~\bibnamefont {Shih}}\ and\ \bibinfo {author} {\bibfnamefont {C.~O.}\ \bibnamefont {Alley}},\ }\href@noop {} {\bibfield  {journal} {\bibinfo  {journal} {Physical Review Letters}\ }\textbf {\bibinfo {volume} {61}},\ \bibinfo {pages} {2921} (\bibinfo {year} {1988})}\BibitemShut {NoStop}%
\bibitem [{\citenamefont {Bouchard}\ \emph {et~al.}(2020)\citenamefont {Bouchard}, \citenamefont {Sit}, \citenamefont {Zhang}, \citenamefont {Fickler}, \citenamefont {Miatto}, \citenamefont {Yao}, \citenamefont {Sciarrino},\ and\ \citenamefont {Karimi}}]{bouchard2020two}%
  \BibitemOpen
  \bibfield  {author} {\bibinfo {author} {\bibfnamefont {F.}~\bibnamefont {Bouchard}}, \bibinfo {author} {\bibfnamefont {A.}~\bibnamefont {Sit}}, \bibinfo {author} {\bibfnamefont {Y.}~\bibnamefont {Zhang}}, \bibinfo {author} {\bibfnamefont {R.}~\bibnamefont {Fickler}}, \bibinfo {author} {\bibfnamefont {F.~M.}\ \bibnamefont {Miatto}}, \bibinfo {author} {\bibfnamefont {Y.}~\bibnamefont {Yao}}, \bibinfo {author} {\bibfnamefont {F.}~\bibnamefont {Sciarrino}}, \ and\ \bibinfo {author} {\bibfnamefont {E.}~\bibnamefont {Karimi}},\ }\href@noop {} {\bibfield  {journal} {\bibinfo  {journal} {Reports on Progress in Physics}\ }\textbf {\bibinfo {volume} {84}},\ \bibinfo {pages} {012402} (\bibinfo {year} {2020})}\BibitemShut {NoStop}%
\bibitem [{\citenamefont {CRAMÉR}(1999)}]{a61aa5fe-d74a-3133-bed2-f35c3c555015}%
  \BibitemOpen
  \bibfield  {author} {\bibinfo {author} {\bibfnamefont {H.}~\bibnamefont {CRAMÉR}},\ }\href {http://www.jstor.org/stable/j.ctt1bpm9r4} {\emph {\bibinfo {title} {Mathematical Methods of Statistics (PMS-9)}}}\ (\bibinfo  {publisher} {Princeton University Press},\ \bibinfo {year} {1999})\BibitemShut {NoStop}%
\bibitem [{\citenamefont {Holevo}(2011)}]{holevo2011probabilistic}%
  \BibitemOpen
  \bibfield  {author} {\bibinfo {author} {\bibfnamefont {A.~S.}\ \bibnamefont {Holevo}},\ }\href@noop {} {\emph {\bibinfo {title} {Probabilistic and statistical aspects of quantum theory}}},\ Vol.~\bibinfo {volume} {1}\ (\bibinfo  {publisher} {Springer Science \& Business Media},\ \bibinfo {year} {2011})\BibitemShut {NoStop}%
\bibitem [{\citenamefont {Rohatgi}\ and\ \citenamefont {Saleh}(2015)}]{rohatgi2015introduction}%
  \BibitemOpen
  \bibfield  {author} {\bibinfo {author} {\bibfnamefont {V.~K.}\ \bibnamefont {Rohatgi}}\ and\ \bibinfo {author} {\bibfnamefont {A.~M.~E.}\ \bibnamefont {Saleh}},\ }\href@noop {} {\emph {\bibinfo {title} {An introduction to probability and statistics}}}\ (\bibinfo  {publisher} {John Wiley \& Sons},\ \bibinfo {year} {2015})\BibitemShut {NoStop}%
\bibitem [{\citenamefont {Knoll}\ \emph {et~al.}(2019)\citenamefont {Knoll}, \citenamefont {Bosyk}, \citenamefont {Grande},\ and\ \citenamefont {Larotonda}}]{knoll2019role}%
  \BibitemOpen
  \bibfield  {author} {\bibinfo {author} {\bibfnamefont {L.~T.}\ \bibnamefont {Knoll}}, \bibinfo {author} {\bibfnamefont {G.~M.}\ \bibnamefont {Bosyk}}, \bibinfo {author} {\bibfnamefont {I.~L.}\ \bibnamefont {Grande}}, \ and\ \bibinfo {author} {\bibfnamefont {M.~A.}\ \bibnamefont {Larotonda}},\ }\href@noop {} {\bibfield  {journal} {\bibinfo  {journal} {Physical Review A}\ }\textbf {\bibinfo {volume} {100}},\ \bibinfo {pages} {062125} (\bibinfo {year} {2019})}\BibitemShut {NoStop}%
\bibitem [{\citenamefont {Knoll}\ and\ \citenamefont {Bosyk}(2023)}]{knoll2023simultaneous}%
  \BibitemOpen
  \bibfield  {author} {\bibinfo {author} {\bibfnamefont {L.~T.}\ \bibnamefont {Knoll}}\ and\ \bibinfo {author} {\bibfnamefont {G.~M.}\ \bibnamefont {Bosyk}},\ }\href@noop {} {\bibfield  {journal} {\bibinfo  {journal} {JOSA B}\ }\textbf {\bibinfo {volume} {40}},\ \bibinfo {pages} {C67} (\bibinfo {year} {2023})}\BibitemShut {NoStop}%
\bibitem [{\citenamefont {Brivio}\ \emph {et~al.}(2010)\citenamefont {Brivio}, \citenamefont {Cialdi}, \citenamefont {Vezzoli}, \citenamefont {Gebrehiwot}, \citenamefont {Genoni}, \citenamefont {Olivares},\ and\ \citenamefont {Paris}}]{brivio2010experimental}%
  \BibitemOpen
  \bibfield  {author} {\bibinfo {author} {\bibfnamefont {D.}~\bibnamefont {Brivio}}, \bibinfo {author} {\bibfnamefont {S.}~\bibnamefont {Cialdi}}, \bibinfo {author} {\bibfnamefont {S.}~\bibnamefont {Vezzoli}}, \bibinfo {author} {\bibfnamefont {B.~T.}\ \bibnamefont {Gebrehiwot}}, \bibinfo {author} {\bibfnamefont {M.~G.}\ \bibnamefont {Genoni}}, \bibinfo {author} {\bibfnamefont {S.}~\bibnamefont {Olivares}}, \ and\ \bibinfo {author} {\bibfnamefont {M.~G.}\ \bibnamefont {Paris}},\ }\href@noop {} {\bibfield  {journal} {\bibinfo  {journal} {Physical Review A}\ }\textbf {\bibinfo {volume} {81}},\ \bibinfo {pages} {012305} (\bibinfo {year} {2010})}\BibitemShut {NoStop}%
\bibitem [{\citenamefont {Tamma}\ and\ \citenamefont {Laibacher}(2023)}]{tamma2023scattershot}%
  \BibitemOpen
  \bibfield  {author} {\bibinfo {author} {\bibfnamefont {V.}~\bibnamefont {Tamma}}\ and\ \bibinfo {author} {\bibfnamefont {S.}~\bibnamefont {Laibacher}},\ }\href@noop {} {\bibfield  {journal} {\bibinfo  {journal} {The European Physical Journal Plus}\ }\textbf {\bibinfo {volume} {138}},\ \bibinfo {pages} {1} (\bibinfo {year} {2023})}\BibitemShut {NoStop}%
\bibitem [{\citenamefont {Laibacher}\ and\ \citenamefont {Tamma}(2015)}]{laibacher2015physics}%
  \BibitemOpen
  \bibfield  {author} {\bibinfo {author} {\bibfnamefont {S.}~\bibnamefont {Laibacher}}\ and\ \bibinfo {author} {\bibfnamefont {V.}~\bibnamefont {Tamma}},\ }\href@noop {} {\bibfield  {journal} {\bibinfo  {journal} {Physical review letters}\ }\textbf {\bibinfo {volume} {115}},\ \bibinfo {pages} {243605} (\bibinfo {year} {2015})}\BibitemShut {NoStop}%
\bibitem [{\citenamefont {Tamma}\ and\ \citenamefont {Laibacher}(2016)}]{tamma2016multi}%
  \BibitemOpen
  \bibfield  {author} {\bibinfo {author} {\bibfnamefont {V.}~\bibnamefont {Tamma}}\ and\ \bibinfo {author} {\bibfnamefont {S.}~\bibnamefont {Laibacher}},\ }\href@noop {} {\bibfield  {journal} {\bibinfo  {journal} {Quantum Information Processing}\ }\textbf {\bibinfo {volume} {15}},\ \bibinfo {pages} {1241} (\bibinfo {year} {2016})}\BibitemShut {NoStop}%
\bibitem [{\citenamefont {Tamma}(2015)}]{tamma2015sampling}%
  \BibitemOpen
  \bibfield  {author} {\bibinfo {author} {\bibfnamefont {V.}~\bibnamefont {Tamma}},\ }\href@noop {} {\bibfield  {journal} {\bibinfo  {journal} {International Journal of Quantum Information}\ }\textbf {\bibinfo {volume} {12}},\ \bibinfo {pages} {1560017} (\bibinfo {year} {2015})}\BibitemShut {NoStop}%
\bibitem [{\citenamefont {Aaronson}\ and\ \citenamefont {Arkhipov}(2011)}]{10.1145/1993636.1993682}%
  \BibitemOpen
  \bibfield  {author} {\bibinfo {author} {\bibfnamefont {S.}~\bibnamefont {Aaronson}}\ and\ \bibinfo {author} {\bibfnamefont {A.}~\bibnamefont {Arkhipov}},\ }in\ \href {\doibase 10.1145/1993636.1993682} {\emph {\bibinfo {booktitle} {Proceedings of the Forty-Third Annual ACM Symposium on Theory of Computing}}},\ \bibinfo {series and number} {STOC '11}\ (\bibinfo  {publisher} {Association for Computing Machinery},\ \bibinfo {address} {New York, NY, USA},\ \bibinfo {year} {2011})\ p.\ \bibinfo {pages} {333–342}\BibitemShut {NoStop}%
\bibitem [{\citenamefont {Tamma}\ and\ \citenamefont {Laibacher}(2021)}]{tamma2021boson}%
  \BibitemOpen
  \bibfield  {author} {\bibinfo {author} {\bibfnamefont {V.}~\bibnamefont {Tamma}}\ and\ \bibinfo {author} {\bibfnamefont {S.}~\bibnamefont {Laibacher}},\ }\href@noop {} {\bibfield  {journal} {\bibinfo  {journal} {Physical Review A}\ }\textbf {\bibinfo {volume} {104}},\ \bibinfo {pages} {032204} (\bibinfo {year} {2021})}\BibitemShut {NoStop}%
\bibitem [{\citenamefont {Razavian}\ \emph {et~al.}(2020)\citenamefont {Razavian}, \citenamefont {Paris},\ and\ \citenamefont {Genoni}}]{e22111197}%
  \BibitemOpen
  \bibfield  {author} {\bibinfo {author} {\bibfnamefont {S.}~\bibnamefont {Razavian}}, \bibinfo {author} {\bibfnamefont {M.~G.~A.}\ \bibnamefont {Paris}}, \ and\ \bibinfo {author} {\bibfnamefont {M.~G.}\ \bibnamefont {Genoni}},\ }\href {\doibase 10.3390/e22111197} {\bibfield  {journal} {\bibinfo  {journal} {Entropy}\ }\textbf {\bibinfo {volume} {22}} (\bibinfo {year} {2020}),\ 10.3390/e22111197}\BibitemShut {NoStop}%
\bibitem [{\citenamefont {Delgado}(2022)}]{sym14091813}%
  \BibitemOpen
  \bibfield  {author} {\bibinfo {author} {\bibfnamefont {F.}~\bibnamefont {Delgado}},\ }\href {\doibase 10.3390/sym14091813} {\bibfield  {journal} {\bibinfo  {journal} {Symmetry}\ }\textbf {\bibinfo {volume} {14}} (\bibinfo {year} {2022}),\ 10.3390/sym14091813}\BibitemShut {NoStop}%
\bibitem [{\citenamefont {Liu}\ \emph {et~al.}(2019)\citenamefont {Liu}, \citenamefont {Yuan}, \citenamefont {Lu},\ and\ \citenamefont {Wang}}]{Liu_2020}%
  \BibitemOpen
  \bibfield  {author} {\bibinfo {author} {\bibfnamefont {J.}~\bibnamefont {Liu}}, \bibinfo {author} {\bibfnamefont {H.}~\bibnamefont {Yuan}}, \bibinfo {author} {\bibfnamefont {X.-M.}\ \bibnamefont {Lu}}, \ and\ \bibinfo {author} {\bibfnamefont {X.}~\bibnamefont {Wang}},\ }\href {\doibase 10.1088/1751-8121/ab5d4d} {\bibfield  {journal} {\bibinfo  {journal} {Journal of Physics A: Mathematical and Theoretical}\ }\textbf {\bibinfo {volume} {53}},\ \bibinfo {pages} {023001} (\bibinfo {year} {2019})}\BibitemShut {NoStop}%
\bibitem [{\citenamefont {Helstrom}(1969)}]{helstrom1969quantum}%
  \BibitemOpen
  \bibfield  {author} {\bibinfo {author} {\bibfnamefont {C.~W.}\ \bibnamefont {Helstrom}},\ }\href@noop {} {\bibfield  {journal} {\bibinfo  {journal} {Journal of Statistical Physics}\ }\textbf {\bibinfo {volume} {1}},\ \bibinfo {pages} {231} (\bibinfo {year} {1969})}\BibitemShut {NoStop}%
\end{thebibliography}%

\newpage
\onecolumngrid
\appendix
\section{Evaluation of the output probabilities in Eq.~\eqref{eq:Probs}}
\label{app:Probs}
The outputs of the scheme are ten and they are represented in TABLE~\ref{demo-table}. In order to evaluate their probabilities, the input state $\ket{\alpha_1}_1\otimes\ket{\alpha_2}_2$ is written in terms of creation operators  (see Eq. \eqref{eq:InitState} for the definition of $\ket{\alpha_1}_1$ and $\ket{\alpha_2}_2$)
\begin{equation}
\ket{\alpha_1}_1\otimes\ket{\alpha_2}_2=\hat{a}^\dagger_{\alpha_1,1}\otimes \hat{a}^\dagger_{\alpha_2,2}\ket{0}\,,
\end{equation}
where
\begin{equation}
\hat{a}^\dagger_{\alpha_i,i}=\cos{\frac{\theta}{2}}\hat{a}^\dagger_{H,i}+\sin{\frac{\theta}{2}}e^{i\phi_i}\hat{a}^\dagger_{V,i}\,\qquad,\qquad i=1,2.\label{aop}
\end{equation}
Here, $\hat{a}^\dagger_{H,i}$ and $\hat{a}^\dagger_{V,i}$ ($i=1,2$) are respectively the creation operator for the horizontal and vertical polarisation state in the $i-$th input channel. The commutation relations are the following $(i,j=1,2)$
\begin{align}
\begin{split}
&\pq{\hat{a}^\dagger_{H,i},\hat{a}^\dagger_{H,j}}=\pq{\hat{a}_{H,i},\hat{a}_{H,j}}=0,\\
&\pq{\hat{a}^\dagger_{V,i},\hat{a}^\dagger_{V,j}}=\pq{\hat{a}_{V,i},\hat{a}_{V,j}}=0,\\
&\pq{\hat{a}_{H,i},\hat{a}^\dagger_{V,j}}=\pq{\hat{a}_{V,i},\hat{a}^\dagger_{H,j}}=0,\\
&\pq{\hat{a}_{H,i},\hat{a}^\dagger_{H,j}}=\pq{\hat{a}_{V,i},\hat{a}^\dagger_{V,j}}=\delta_{i,j}.
\end{split}\label{comm}
\end{align}
In this way it is proven that the commutator of any linear combination of these operators is, in fact, a number. It is possible to take as an example
\begin{equation}
\pq{\hat{a}_{\alpha_i,k},\hat{a}^\dagger_{\alpha_j,l}}=\pt{\cos^2{\frac{\theta}{2}}+\sin^2{\frac{\theta}{2}}e^{i\pt{\phi_j-\phi_i}}}\delta_{k,l}\label{B4}
\end{equation}
This result is also equal to $\bra{0}\hat{a}_{\alpha_i,k}\hat{a}^\dagger_{\alpha_j,l}\ket{0}$. In fact, since the commutator is a number, one can write
\begin{equation}
\bra{0}\hat{a}_{\alpha_i,k}\hat{a}^\dagger_{\alpha_j,l}\ket{0}=\bra{0}\pq{\hat{a}_{\alpha_i,k},\hat{a}^\dagger_{\alpha_j,l}}\ket{0}+\bra{0}\hat{a}^\dagger_{\alpha_j,l}\hat{a}_{\alpha_i,k}\ket{0}=\pq{\hat{a}_{\alpha_i,k},\hat{a}^\dagger_{\alpha_j,l}}\braket{0}{0}=\pq{\hat{a}_{\alpha_i,k},\hat{a}^\dagger_{\alpha_j,l}}.\label{braketcomm}
\end{equation}
The beam splitter can be described as a unitary matrix with transition amplitudes
\begin{equation}
U_{BS}=\frac{1}{\sqrt{2}}\begin{pmatrix}
1&1\\
-1&1
\end{pmatrix}
\end{equation}
and it acts on the injected probe through the map
\begin{equation}
\hat{U}_{BS}\hat{a}^\dagger_{X,i}\hat{U}_{BS}^\dagger = \sum_{j=1,2}\pt{U_{BS}}_{i,j}\hat{a}^\dagger_{X,j},\qquad X=H,V
\end{equation}
The two-photon state at the output of the BS thus reads
\begin{align}
U_{BS}\ket{\psi_{in}}&=\frac{1}{2}\pt{\hat{a}^\dagger_{\alpha_1,1}\otimes I_2+I_1\otimes \hat{a}^\dagger_{\alpha_1,2}}\pt{-\hat{a}^\dagger_{\alpha_2,1}\otimes I_2+I_1\otimes \hat{a}^\dagger_{\alpha_2,2}}\ket{0}=\nonumber\\
&=\frac{1}{2}\pt{-\hat{a}^\dagger_{\alpha_1,1}\hat{a}^\dagger_{\alpha_2,1}\otimes I_2-\hat{a}^\dagger_{\alpha_2,1}\otimes \hat{a}^\dagger_{\alpha_1,2}+\hat{a}^\dagger_{\alpha_1,1}\otimes \hat{a}^\dagger_{\alpha_2,2}+I_1\otimes \hat{a}^\dagger_{\alpha_1,2}\hat{a}^\dagger_{\alpha_2,2}}\ket{0}.\label{Upsi}
\end{align}
The new operators of creation refer to the output channels 1 and 2. In these channels, the photons are projected by the PBSs into two bases, namely $\pg{\ket{H}_i,\ket{V}_i}$, with $i=1,2$.

\begin{table}
\begin{tabular}{l|l|l|llll|}
\cline{2-7}
                                         & Output state & Probability & $D_{H1}$ & $D_{V1}$ & $D_{H2}$ & $D_{V2}$ \\ \hline
\multicolumn{1}{|l|}{\multirow{2}{*}{\begin{turn}{90}$DB_H$\end{turn}}} &  $a^\dagger_{H,1}a^\dagger_{H,1}\otimes I_2\ket{0}$            & $P_{H,1;H,1}=\frac{1}{8}\pt{1+\cos\theta}^2$           & \multicolumn{1}{l|}{\checkmark} & \multicolumn{1}{l|}{}  & \multicolumn{1}{l|}{}  &   \\ \cline{2-7} 
\multicolumn{1}{|l|}{}              & $I_1\otimes a^\dagger_{H,2}a^\dagger_{H,2}\ket{0}$            & $P_{H,2;H,2}=\frac{1}{8}\pt{1+\cos\theta}^2$           &       \multicolumn{1}{l|}{}  & \multicolumn{1}{l|}{} & \multicolumn{1}{l|}{\checkmark}  &   \\ \hline 
\multicolumn{1}{|l|}{\multirow{2}{*}{\begin{turn}{90}$DB_V$\end{turn}}}        & $a^\dagger_{V,1}a^\dagger_{V,1}\otimes I_2\ket{0}$            & $P_{V,1;V,1}=\frac{1}{8}\pt{1-\cos\theta}^2$          &           \multicolumn{1}{l|}{}  & \multicolumn{1}{l|}{\checkmark}  & \multicolumn{1}{l|}{} &   \\ \cline{2-7} 
\multicolumn{1}{|l|}{}                   & $I_2\otimes a^\dagger_{V,2}a^\dagger_{V,2}\ket{0}$            & $P_{V,2;V,2}=\frac{1}{8}\pt{1-\cos\theta}^2$           & \multicolumn{1}{l|}{}  & \multicolumn{1}{l|}{}  & \multicolumn{1}{l|}{}  & \checkmark \\ \hline
\multicolumn{1}{|l|}{\multirow{2}{*}{\begin{turn}{90}$SB$\end{turn}}} & $a^\dagger_{H,1}a^\dagger_{V,1}\otimes I_2\ket{0}$            & $P_{H,1;V,1}=\frac{1}{4}\sin^2\theta\cos^2\delta_\phi$           & \multicolumn{1}{l|}{\checkmark} & \multicolumn{1}{l|}{\checkmark} & \multicolumn{1}{l|}{}  &   \\ \cline{2-7} 
\multicolumn{1}{|l|}{}                   & $I_2\otimes a^\dagger_{H,2}a^\dagger_{V,2}\ket{0}$            & $P_{H,2;V,2}=\frac{1}{4}\sin^2\theta\cos^2\delta_\phi$           & \multicolumn{1}{l|}{}  & \multicolumn{1}{l|}{}  & \multicolumn{1}{l|}{\checkmark} & \checkmark \\ \hline
\multicolumn{1}{|l|}{\multirow{4}{*}{\begin{turn}{90}$C$\end{turn}}} & $a^\dagger_{H,1}\otimes a^\dagger_{H,2}\ket{0}$            & $P_{H,1;H,2}=0$           & \multicolumn{1}{l|}{\checkmark} & \multicolumn{1}{l|}{}  & \multicolumn{1}{l|}{\checkmark} &   \\ \cline{2-7} 
\multicolumn{1}{|l|}{}                   & $a^\dagger_{V,1}\otimes a^\dagger_{V,2}\ket{0}$            & $P_{V,1;V,2}=0$           & \multicolumn{1}{l|}{}  & \multicolumn{1}{l|}{\checkmark} & \multicolumn{1}{l|}{}  & \checkmark \\ \cline{2-7} 
\multicolumn{1}{|l|}{}                   & $a^\dagger_{H,1}\otimes a^\dagger_{V,2}\ket{0}$            & $P_{H,1;V,2}=\frac{1}{4}\sin^2\theta\sin^2\delta_\phi$           & \multicolumn{1}{l|}{\checkmark} & \multicolumn{1}{l|}{}  & \multicolumn{1}{l|}{}  & \checkmark \\ \cline{2-7} 
\multicolumn{1}{|l|}{}                   & $a^\dagger_{V,1}\otimes a^\dagger_{H,2}\ket{0}$            & $P_{V,1;H,2}=\frac{1}{4}\sin^2\theta\sin^2\delta_\phi$           & \multicolumn{1}{l|}{}  & \multicolumn{1}{l|}{\checkmark} & \multicolumn{1}{l|}{\checkmark} &   \\ \hline
\end{tabular}
\caption{\label{demo-table}Possible outcome states (first column) with their respective probabilities (second column). The last four columns represent the detectors that clicks for each event. The events are divided in double bunching events ($DB_H$ and $DB_V$), single bunching events ($SB$) and coincidence events ($C$).}
\end{table}

\subsubsection*{Coincidence}
Using Eqs.~\eqref{braketcomm} and~\eqref{Upsi}, the probability $P_{X,1;Y,2}$ to observe the two photons in different output channels and with polarisation $X,Y=H,V$, namely in the state
\begin{equation}
 \ket{X}_1\otimes\ket{Y}_2 =  \hat{a}_{X,1}^\dagger\hat{a}_{Y,2}^\dagger\ket{0}
\end{equation}
 is
\begin{align}
P_{X,1;Y,2}&=\left\vert \bra{0}\hat{a}_{X,1}\hat{a}_{Y,2}\hat{U}_{BS}\hat{a}^\dagger_{\alpha_1,1}\hat{a}^\dagger_{\alpha_2,2}\ket{0}\right\vert^2=\frac{1}{4}\left\vert \bra{0}\hat{a}_{X,1}\hat{a}_{Y,2}\pt{\hat{a}^\dagger_{\alpha_1,1}\hat{a}^\dagger_{\alpha_2,2}-\hat{a}^\dagger_{\alpha_1,2}\hat{a}^\dagger_{\alpha_2,1}}\ket{0}            \right\vert^2=\\
&=\frac{1}{4}\Big\vert \pq{\hat{a}_{X,1},\hat{a}^\dagger_{\alpha_1,1}}\pq{\hat{a}_{Y,2},\hat{a}^\dagger_{\alpha_2,2}}+\pq{\hat{a}_{X,1},\hat{a}^\dagger_{\alpha_2,2}}\pq{\hat{a}_{Y,2},\hat{a}^\dagger_{\alpha_1,1}}-\\&-\pq{\hat{a}_{X,1},\hat{a}^\dagger_{\alpha_1,2}}\pq{\hat{a}_{Y,2},\hat{a}^\dagger_{\alpha_2,1}} -\pq{\hat{a}_{X,1},\hat{a}^\dagger_{\alpha_2,1}}\pq{\hat{a}_{Y,2},\hat{a}^\dagger_{\alpha_1,2}}           \Big\vert^2.
\end{align}
Using the definition in Eq.~\eqref{aop} and the commutation relations in Eqs.~\eqref{comm}, it is possible to write
\begin{align}
P_{X,1;Y,2}&=\frac{1}{4}\Big\vert \pq{\hat{a}_{X,1},\hat{a}^\dagger_{\alpha_1,1}}\pq{\hat{a}_{Y,2},\hat{a}^\dagger_{\alpha_2,2}}-\pq{\hat{a}_{X,1},\hat{a}^\dagger_{\alpha_2,1}}\pq{\hat{a}_{Y,2},\hat{a}^\dagger_{\alpha_1,2}}           \Big\vert^2,
\end{align}
with the following results by using Eq~\eqref{B4}
\begin{align}
   &P_{H,1;H,2}=P_{V,1;V,2}=0, \\
   &P_{H,1;V,2}=P_{V,1;H,2}=\frac{1}{4}\sin^2\theta\sin^2\delta_\phi,
\end{align}
where
\begin{align}
    \delta_\phi=\left\vert\frac{\phi_1-\phi_2}{2}\right\vert.
    \label{parameters}
\end{align}
This happens because the coincidence event cannot be observed if the two output photons are identical. Also, $P_{V,1;V,2}=P_{V,1;V,2}$ because of the symmetry of the setup. In fact, both photons have the same probability of being reflected or transmitted from the BS, and the PBS projects into the same base.
Since these two events have identical probabilities to be observed, resolving them separately does not yield more information than considering a unique, global coincidence event with probability
\begin{equation}
    P_C=P_{H,1;V,2}+P_{V,1;H,2}=\frac{1}{2}\sin^2\theta\sin^2\delta_\phi.
    \label{eq:PCApp}
\end{equation}
In fact, evaluating the term of the Fisher information associated to a general probability $P=2P'$ it is possible to obtain
\begin{align}
    F_P&=P\nabla\nabla\log P=2P'\nabla\nabla\log 2P'=2P'\nabla\nabla\log P'=2F_{P'},
\end{align}
where $X,Y=\theta,\delta_\phi$.
\subsubsection*{Double Bunching} 
One can evaluate the probability $P_{X,1;X,1}$ to detect the two photons with the same polarisation $X=H,V$ and in the same output channel $1$, i.e. in the output state
\begin{equation}
    \ket{X}_1\otimes\ket{X}_1=\frac{\hat{a}_{X,1}^\dagger\hat{a}_{X,1}^\dagger}{\sqrt{2}}\ket{0}
\end{equation}
from Eq.~\eqref{Upsi}
\begin{align}
P_{X,1;X,1}&=\left\vert \bra{0}\frac{\hat{a}_{X,1}\hat{a}_{X,1}}{\sqrt{2}}\hat{U}_{BS}\hat{a}^\dagger_{\alpha_1,1}\hat{a}^\dagger_{\alpha_2,2}\ket{0}\right\vert^2=\frac{1}{4}\left\vert \frac{1}{\sqrt{2}}\bra{0}\hat{a}_{X,1}\hat{a}_{X,1}\hat{a}^\dagger_{\alpha_1,1}\hat{a}^\dagger_{\alpha_2,1}\ket{0}\right\vert^2=\\
&=\frac{1}{8}\left\vert\bra{0}\pt{\hat{a}_{X,1}\pq{\hat{a}_{X,1},\hat{a}^\dagger_{\alpha_1,1}}\hat{a}^\dagger_{\alpha_2,1}+\pq{\hat{a}_{X,1},\hat{a}^\dagger_{\alpha_1,1}}\hat{a}_{X,1}\hat{a}^\dagger_{\alpha_2,1}+\hat{a}^\dagger_{\alpha_1,1}\hat{a}_{X,1} \hat{a}_{X,1}\hat{a}^\dagger_{\alpha_2,1}}\ket{0}\right\vert^2=\nonumber\\
&=\frac{1}{8}\left\vert 2\bra{0}\pq{\hat{a}_{X,1},\hat{a}^\dagger_{\alpha_1,1}}\hat{a}_{X,1}\hat{a}^\dagger_{\alpha_2,1}\ket{0}\right\vert^2=\frac{1}{2}\left\vert\pq{\hat{a}_{X,1},\hat{a}^\dagger_{\alpha_1,1}}\right\vert^2\left\vert \pq{\hat{a}_{X,1},\hat{a}^\dagger_{\alpha_2,1}}\right\vert^2.
\label{Paa}
\end{align}
Using the definition in Eq.~\eqref{aop} and the commutation relations in Eqs.~\eqref{comm}, it is possible to write
\begin{align}
&P_{H,1;H,1}=\frac{1}{2}\cos^4\frac{\theta}{2}=\frac{1}{8}\pt{1+\cos\theta}^2,\\
&P_{V,1;V,1}=\frac{1}{2}\sin^4\frac{\theta}{2}=\frac{1}{8}\pt{1-\cos\theta}^2.
\end{align}
For the symmetry of the setup, it is possible to prove that
\begin{equation}
   P_{H,1;H,1}=P_{H,2;H,2}\qquad , \qquad  P_{V,1;V,1}= P_{V,2;V,2}.
\end{equation}
Similarly to the case of coincidence events probability, this allows one to not resolve all four events separately, and instead to classify the single bunching events as $DB_H$ and $DB_V$ such as
\begin{equation}
    P_{DB_H}=P_{H,1;H,1}+P_{H,2;H,2}=\frac{1}{4}\pt{1+\cos\theta}^2,
    \label{eq:PDBHApp}
\end{equation}
\begin{equation}
    P_{DB_V}=P_{V,1;V,1}+P_{V,2;V,2}=\frac{1}{4}\pt{1-\cos\theta}^2.
    \label{eq:PDBVApp}
\end{equation}
\subsubsection*{Single bunching}
Using Eq.~\eqref{Upsi}, the probability $P_{H,1;V,1}$ to observe the two photons in the same output channel and with different polarisations, i.e. in the output state
\begin{equation}
    \ket{H}_1\otimes\ket{V}_1=\hat{a}_{H,1}^\dagger\hat{a}_{V,1}^\dagger\ket{0}
\end{equation}
reads (using the definition in Eq.~\eqref{parameters})
\begin{align}
P_{H,1;V,1}&=\left\vert \bra{0}\hat{a}_{H,1}\hat{a}_{V,1}\hat{U}_{BS}\hat{a}^\dagger_{\alpha_1,1}\hat{a}^\dagger_{\alpha_2,2}\ket{0}\right\vert^2=\frac{1}{4}\left\vert \bra{0}\hat{a}_{H,1}\hat{a}_{V,1}\hat{a}^\dagger_{\alpha_1,1}\hat{a}^\dagger_{\alpha_2,1}\ket{0}\right\vert^2\nonumber\\
&=\frac{1}{4}\left\vert \pq{\hat{a}_{H,1},\hat{a}^\dagger_{\alpha_1,1}}\pq{\hat{a}_{V,1},\hat{a}^\dagger_{\alpha_2,1}}+\pq{\hat{a}_{H,1},\hat{a}^\dagger_{\alpha_2,1}}\pq{\hat{a}_{V,1},\hat{a}^\dagger_{\alpha_1,1}} \right\vert^2=\\
&=\frac{1}{4}\left\vert e^{i\phi_2}\cos\frac{\theta}{2}\sin\frac{\theta}{2}+e^{i\phi_1}\cos\frac{\theta}{2}\sin\frac{\theta}{2} \right\vert^2=\frac{1}{4}\sin^2\theta\cos^2\delta_\phi.
\end{align}
For the symmetry of the scheme, $P_{H,1;V,1}=P_{H,2;V,2}$. Therefore, it is once again possible to define only one event, the single bunching SB event, with probability
\begin{equation}
    P_{SB}=P_{H,1;V,1}+P_{H,2;V,2}=\frac{1}{2}\sin^2\theta\cos^2\delta_\phi.
    \label{eq:PSBApp}
\end{equation}

Eqs.~\eqref{eq:PCApp},~\eqref{eq:PDBHApp},~\eqref{eq:PDBVApp},~\eqref{eq:PSBApp} coincide with the expression of the probabilities in Eq.~\eqref{eq:Probs} in the main text.

\section{Maximum likelihood estimators}
\label{app:Likelihood}
Using the probabilities defined in TABLE~\ref{demo-table}, the log-likelihood function after $N$ repetition of the experiment can be calculated as
\begin{equation}
\log\mathcal{L}\pt{\theta,\delta_\phi}=\log P_{DB_H}^{N_{DB_H}}P_{DB_V}^{N_{DB_V}}P_{SB}^{N_{SB}}P_{C}^{N_{C}},
\end{equation}
where $N_{DB_H}$, $N_{DB_V}$, $N_{SB}$ and $N_C$ are the numbers of registered double-bunching events where both photon have polarisation $H$, of double-bunching events where both photon have polarisation $V$, of single bunching events and coincidence events respectively, with $N_{DB_H}+N_{DB_V}+N_{SB}+N_C=N$.
The equations for the maximum-likelihood estimators
\begin{equation}
\begin{cases}
\frac{\partial}{\partial \theta}\log\mathcal{L}\pt{\theta,\delta_\phi}=0\Rightarrow\frac{2N}{\sin\tilde{\theta}}\pt{\cos\tilde\theta-\frac{N_{DB_H}-N_{DB_V}}{N}}=0
\\
\frac{\partial}{\partial \delta_\phi}\log\mathcal{L}\pt{\theta,\delta_\phi}=0\Rightarrow\frac{2N_{SB}}{\tan\tilde{\delta}_\phi}\pt{\frac{N_C}{N_{SB}}-\tan^2\tilde{\delta}_\phi}=0
\end{cases},
\end{equation}
yield to a single stationary point $\pt{\tilde{\theta},\tilde{\delta}_\phi}$ with
\begin{equation}
\begin{cases}
\tilde{\theta}=\arccos\frac{N_{DB_H}-N_{DB_V}}{N}\\
\tilde{\delta}_\phi=\arctan\sqrt{\frac{N_C}{N_{SB}}}
\end{cases}.
\label{eq:MLEApp}
\end{equation}
In order to see if this point is a point of maximum it is necessary to evaluate the Hessian. It is easy to show that the Hessian is diagonal, and the non-zero elements evaluated at $\pt{\tilde{\theta},\tilde{\delta}_\phi}$ are
\begin{equation}
    \begin{cases}
      \frac{\partial^2}{\partial \theta^2}\log\mathcal{L}\pt{\theta,\delta_\phi}=-\frac{2N}{\sin^2\tilde{\theta}}\pt{1+\cos\tilde{\theta}\frac{N_{DB_H}-N_{DB_V}}{N}}=-2N\frac{N^2+\pt{N_{DB_H}-N_{DB_V}}^2}{N^2-\pt{N_{DB_H}-N_{DB_V}}^2} < 0 \\
      \frac{\partial^2}{\partial \delta^2_\phi}\log\mathcal{L}\pt{\theta,\delta_\phi}=-\frac{2N_{SB}}{\sin^2\tilde{\delta}_\phi}\pt{\tan^2\tilde{\delta}_\phi+\frac{N_{C}}{N_{SB}}}=-4\pt{N_C+N_{SB}}< 0
    \end{cases},
\end{equation}
for $N_C+N_{SB}>0$, which proves that the point $\pt{\tilde{\theta},\tilde{\delta}_\phi}$ is a point of maximum of the likelihood and that the quantities $\tilde{\theta}$ and $\tilde{\delta}_\phi$ are the maximum-likelihood estimators.
It can be noticed that, for every fixed value of $N$, the estimators $\tilde{\theta}$ and $\tilde{\delta}_\phi$ can take only a finite number of possible values.

\section{Evaluation of the quantum Fisher information matrix in Eq.~\eqref{QFIMi}}
\label{app:A}

It is well known in literature that the QFIM for the parameters $\pt{\theta, \phi_i}$ of a pure polarization state, as the ones in Eq.\eqref{eq:InitState}, can be written as~\cite{e22111197,sym14091813,Liu_2020}
\begin{equation}
H_{\ket{\alpha_i}_i}\pt{\theta, \phi_i}=\begin{pmatrix}
1&0\\
0&\sin^2\theta
\end{pmatrix}.\label{QFIMThe}
\end{equation}

Since the input state is defined by the tensor product of two single-photon states, the QFIM associated with the estimation of $\pt{\theta, \phi_1, \phi_2}$ immediately derives from the single-photon QFIM~\eqref{QFIMThe} of the two states~\cite{e22111197,Liu_2020}
\begin{equation}
H\pt{\theta, \phi_1, \phi_2}
%=H_{\ket{\alpha_1}_1}\pt{\theta, \phi_1}+H_{\ket{\alpha_2}_2}\pt{\theta, \phi_2}
=\begin{pmatrix}
2&0&0\\
0&\sin^2\theta&0\\
0&0&\sin^2\theta
\end{pmatrix}.\label{GeneralH}
\end{equation}
This is a well known property of the QFIM: the information achievable from two different (and disentangled) states is additive.

Defining
\begin{equation}
m_\phi=\frac{\phi_1+\phi_2}{2}\qquad ,\qquad\delta_\phi=\left\vert\frac{\phi_1-\phi_2}{2}\right\vert,
\end{equation}
it is possible to write a Jacobian for the transformation from the parameters $\pt{\theta,\phi_1,\phi_2}$ to $\pt{\theta,m_\phi,\delta_\phi}$
\begin{equation}
    J=\frac{1}{2}\begin{pmatrix}
        2&0&0\\
        0&\mathrm{sgn}\pt{\phi_1-\phi_2}&\mathrm{sgn}\pt{\phi_1-\phi_2}\\
        0&\mathrm{sgn}\pt{\phi_1-\phi_2}&-\mathrm{sgn}\pt{\phi_1-\phi_2}
    \end{pmatrix}\qquad , \qquad J^{-1}=\begin{pmatrix}
        1&0&0\\
        0&\mathrm{sgn}\pt{\phi_1-\phi_2}&\mathrm{sgn}\pt{\phi_1-\phi_2}\\
        0&\mathrm{sgn}\pt{\phi_1-\phi_2}&-\mathrm{sgn}\pt{\phi_1-\phi_2}
    \end{pmatrix}.
\end{equation}
$H\pt{\theta_1,m_\phi,\delta_\phi}$ is linked to $H\pt{\theta, \phi_1, \phi_2}$ by the transformation
\begin{equation}
H\pt{\theta_1, m_\phi,\delta_\phi}=\pt{J^{-1}}^TH\pt{\theta, \phi_1, \phi_2}J^{-1}=2\begin{pmatrix}
1&0&0\\
0&\sin^2\theta&0\\
0&0&\sin^2\theta
\end{pmatrix}\label{transform}.
\end{equation}
So, since the QFI is diagonal, it is possible to evaluate the QFI for the estimation of the parameters $\pt{\theta,\delta_\phi}$ by taking the submatrix
\begin{equation}
H\pt{\theta, \delta_\phi}=2\begin{pmatrix}
1&0\\
0&\sin^2\theta
\end{pmatrix}.
\end{equation}

\section{Fisher information matrix in Eq.~\eqref{QFIMi}}
\label{app:Fisher}
Employing the definition of the FIM~\cite{a61aa5fe-d74a-3133-bed2-f35c3c555015,rohatgi2015introduction, helstrom1969quantum, holevo2011probabilistic}, it is evaluated as the sum of four different terms, each one associated with the contribution of one of the outcome events probabilities $P_{DB_H}$, $P_{DB_V}$, $P_{SB}$, $P_{C}$ in Eqs.~\eqref{eq:Probs}, i.e.
\begin{equation}
F\pt{\theta,\delta_\phi}=F_{DB_H}\pt{\theta,\delta_\phi}+F_{DB_V}\pt{\theta,\delta_\phi}+F_{SB}\pt{\theta,\delta_\phi}+F_C\pt{\theta,\delta_\phi},\label{defFIM}
\end{equation}
where
\begin{equation}
    F_X=\frac{1}{P_X}\begin{pmatrix}
        \pt{\partial_{\theta}P_{X}}^2 & \partial_{\theta}P_{X}\partial_{\delta_\phi}P_{X}\\
\partial_{\theta}P_{X}\partial_{\delta_\phi}P_{X} & \pt{\partial_{\delta_\phi}P_{X}}^2
    \end{pmatrix}\qquad , \qquad X=DB_H, DB_V, SB, C.
\end{equation}
By using the probabilities in Eqs~\eqref{eq:PCApp},\eqref{eq:PDBHApp},\eqref{eq:PDBVApp} and~\eqref{eq:PSBApp} we found
\begin{align}
F_{DB_H}\pt{\theta,\delta_\phi}&=\frac{1}{P_{DB_H}}\begin{pmatrix}
\pt{\partial_{\theta}P_{DB_H}}^2 & \partial_{\theta}P_{DB_H}\partial_{\delta_\phi}P_{DB_H}\\
\partial_{\theta}P_{DB_H}\partial_{\delta_\phi}P_{DB_H} & \pt{\partial_{\delta_\phi}P_{DB_H}}^2
\end{pmatrix}=\\&=\begin{pmatrix}
\sin^2 \theta&0\\
0&0
\end{pmatrix},
\end{align}
\begin{align}
F_{DB_V}\pt{\theta,\delta_\phi}&=\frac{1}{P_{DB_V}}\begin{pmatrix}
\pt{\partial_{\theta}P_{DB_V}}^2 & \partial_{\theta}P_{DB_V}\partial_{\delta_\phi}P_{DB_V}\\
\partial_{\theta}P_{DB_V}\partial_{\delta_\phi}P_{DB_V} & \pt{\partial_{\delta_\phi}P_{DB_V}}^2
\end{pmatrix}=\\&=\begin{pmatrix}
\sin^2 \theta&0\\
0&0
\end{pmatrix},
\end{align}
\begin{align}
F_{SB}\pt{\theta,\delta_\phi}&=\frac{1}{P_{SB}}\begin{pmatrix}
\pt{\partial_{\theta}P_{SB}}^2 & \partial_{\theta}P_{SB}\partial_{\delta_\phi}P_{SB}\\
\partial_{\theta}P_{SB}\partial_{\delta_\phi}P_{SB} & \pt{\partial_{\delta_\phi}P_{SB}}^2
\end{pmatrix}=\\&=2\begin{pmatrix}
\cos^2 \theta \cos^2\delta_\phi&-\frac{1}{4}\sin 2\theta\sin 2\delta_\phi\\
-\frac{1}{4}\sin 2\theta\sin 2\delta_\phi&\sin^2 \theta\sin^2\delta_\phi
\end{pmatrix},
\end{align}
\begin{align}
F_{C}\pt{\theta,\delta_\phi}&=\frac{1}{P_{C}}\begin{pmatrix}
\pt{\partial_{\theta}P_{C}}^2 & \partial_{\theta}P_{C}\partial_{\delta_\phi}P_{C}\\
\partial_{\theta}P_{SB}\partial_{\delta\phi2}P_{C} & \pt{\partial_{\delta_\phi}P_{C}}^2
\end{pmatrix}=\\&=2\begin{pmatrix}
\cos^2 \theta \sin^2\delta_\phi&\frac{1}{4}\sin 2\theta\sin 2\delta_\phi\\
\frac{1}{4}\sin 2\theta\sin 2\delta_\phi&\sin^2 \theta\cos^2\delta_\phi
\end{pmatrix}.
\end{align}
Summing all the terms Eq~\ref{defFIM} reduces to
\begin{equation}
F\pt{\theta,\delta_\phi}=\begin{pmatrix}
2&0\\0&2\sin^2 \theta
\end{pmatrix}.
\end{equation} 

\section{Probability distributions of the maximum likelihood estimators}
\label{app:Prob}

It is now possible to evaluate the probability distributions of the maximum-likelihood estimators $\tilde{\theta}$ and $\tilde{\delta}_\phi$ in Eq.~\eqref{eq:MLEApp}, for fixed values of the total number of repetitions of the experiment $N$ of sampling measurements and the true unknown values of the polarisation parameters $\theta$ and $\delta_\phi$. 
For example, the probability of having a certain $N_{DB_H}$ and a certain $N_{DB_V}$ given $N$ repetitions of the experiment can be expressed as the multinomial probability
\begin{align}
P\pt{N_{DB_H},N_{DB_V}|N}&=\frac{N!}{N_{DB_H}!N_{DB_V}!\pt{N_C+N_{SB}}!}P_{DB_H}^{N_{DB_H}}P_{DB_V}^{N_{DB_V}}\pt{P_C+P_{SB}}^{N_C+N_{SB}}.
\end{align}
However, in general, multiple pairs $\left(N_{DB_H},N_{DB_V}\right)$ yield the same value of $\tilde{\theta}$.
Using the expression of $\tilde{\theta}$ in Eq.~\eqref{eq:MLEApp} it is possible to write the probability of having a certain $\tilde{\theta}$ and $N_{DB_V}$ for a fixed $N$, as
\begin{align}
P\pt{\tilde{\theta},N_{DB_V}|N}&=\frac{N!}{\left(N\cos\tilde{\theta}+N_{DB_V}\right)!N_{DB_V}!\left(N\left(1-\cos\tilde{\theta}\right)-2N_{DB_V}\right)!}\times\\&\times P_{DB_H}^{N\cos\tilde{\theta}+N_{DB_V}}P_{DB_V}^{N_{DB_V}}\pt{P_C+P_{SB}}^{N\pt{1-\cos\tilde{\theta}}-2N_{DB_V}},
\end{align}
which, specifying the expressions of the probabilities in Eq.~\eqref{eq:Probs}, becomes
\begin{align}
P\pt{\tilde{\theta},N_{DB_V}|N}&=\pt{\cos^2\frac{\theta}{2}}^{2N\cos^2\frac{\tilde{\theta}}{2}}\pt{\sin^2\frac{\theta}{2}}^{2N\sin^2\frac{\tilde{\theta}}{2}}\times\\
&\times\frac{N!}{\left(N\cos\tilde{\theta}+N_{DB_V}\right)!N_{DB_V}!\left(N\left(1-\cos\tilde{\theta}\right)-2N_{DB_V}\right)!}2^{N\pt{1-\cos\tilde{\theta}}-2N_{DB_V}}.
\end{align}
Then, the probability of having $\tilde{\theta}$ fixing $N$ is the sum of the $P\pt{\tilde{\theta},N_{DB_V}|N}$ with all the possible $N_{DB_V}$. This leads to a marginal distribution, that has been obtained from a multinomial distribution. The result will be a binomial distribution
\begin{align}
P\pt{\tilde{\theta}|N}&=\sum_{N_{BD_V}}P\pt{\tilde{\theta},N_{DB_V}|N}=\\
&=\frac{\pt{2N}!}{\pt{2N\cos^2\frac{\tilde{\theta}}{2}}!\pt{2N\sin^2\frac{\tilde{\theta}}{2}}!}\pt{\cos^2\frac{\theta}{2}}^{2N\cos^2\frac{\tilde{\theta}}{2}}\pt{\sin^2\frac{\theta}{2}}^{2N\sin^2\frac{\tilde{\theta}}{2}}.
\end{align}
The cumulative distribution for the parameter $\tilde{\theta}$ with $N$ fixed is given by 
\begin{equation}
C_N\pt{\tilde{\theta}_{\mathrm{tr}}}=P\pt{\tilde{\theta}\leq\tilde{\theta}_{\mathrm{tr}}\vert N}=\sum_{\tilde{\theta=-1}}^{\tilde{\theta}_{\mathrm{tr}}}P\pt{\tilde{\theta}|N}\label{CumDefinTh}
\end{equation}
and defines the probability of having a value of $\tilde{\theta}$ inferior to a threshold $\tilde{\theta}_{\mathrm{tr}}$.

With an analogous procedure, it is possible to find also $P\pt{\tilde{\delta}_\phi|N}$. 
Since $\tilde{\delta}_\phi$ depends only to $N_C$ and $N_{SB}$, it is necessary first to find $P\pt{N_C,N_{SB}|N}$, that is a multinomial
\begin{align}
P\pt{N_{C},N_{SB}|N}&=\frac{N!}{N_{C}!N_{SB}!\pt{N_{DB_H}+N_{DB_V}}!}P_{C}^{N_{C}}P_{SB}^{N_{SB}}\pt{P_{DB_H}+P_{DB_V}}^{N_{DB_H}+N_{DB_V}}.
\end{align}
Then, using the formula for $\tilde{\delta}_\phi$, it is possible to write
\begin{align}
P\pt{\tilde{\delta}_\phi,N_{SB}|N}&=\frac{N!}{\pt{N_{SB}\tan^2\tilde{\delta}_\phi}!N_{SB}!\pt{N-N_{SB}\pt{1+\tan^2\tilde{\delta}_\phi}}!}\times\\
&\times P_{C}^{N_{SB}\tan^2\tilde{\delta}_\phi}P_{SB}^{N_{SB}}\pt{P_{DB_H}+P_{DB_V}}^{N-N_{SB}\pt{1+\tan^2\tilde{\delta}_\phi}}.
\end{align}
The probability of having $\tilde{\delta}_\phi$ as a result of an estimation using a sample with $N$ elements, is the sum of the $P\pt{\tilde{\delta}_\phi,N_{SB}|N}$ with all the possible value of $N_{SB}$
\begin{equation}
P\pt{\tilde{\delta}_\phi|N}=\sum_{N_{SB}}P\pt{\tilde{\delta}_\phi,N_{SB}|N}.
\label{eq:ProbDeltaApp}
\end{equation}

It is important to notice that the estimator $\tilde{\theta}$ in Eq.~\eqref{eq:MLEApp} is well defined for every possible outcome $(N_{DB_H},N_{DB_V},N_{SB},N_{C})$. 
Instead, $\tilde{\delta}_\phi$ is indeterminate if $N_{SB}=N_{C}=0$. 
The probability that the outcome of the experiment after one trial is not SB or C is $P_{DB_H}+P_{DB_V}$. After $N$ repetition of the experiment, the probability that $N_{SB}=N_{C}=0$ is
\begin{equation}
P_{\mathrm{fail}}=\pt{P_{DB_H}+P_{DB_V}}^N=\pt{\frac{1+\cos^2\theta}{2}}^N.
\end{equation}
This means that when $\theta=0,\pi$ it is not possible to retrieve $\delta_\phi$. This is in accord to the fact that when $\theta=0,\pi$, the state of the input photons in Eq.~\eqref{eq:InitState} does not depend on $\phi_1$ and $\phi_2$, and ultimately on $\delta_\phi$. 
Moreover, the further $\theta$ is from $0,\pi$, the smaller is $P_{\mathrm{fail}}$. Also, when $\theta\neq 0,\pi$, increasing the number $N$ helps in reducing $P_{\mathrm{fail}}$. 
As a consequence, the probability $P\pt{\tilde{\delta}_\phi|N}$, as defined in Eq.~\eqref{eq:ProbDeltaApp}, sums to $1-P_{\mathrm{fail}}$ instead of $1$. 
However, it is always possible to redefine $P\pt{\tilde{\delta}_\phi|N}$ as the probability distribution of $\tilde{\delta}_\phi$ conditioned to observing a well defined value of $\tilde{\delta}_\phi$. In other words it is possible to discard the failure events and normalize the probabilities
\begin{equation}
P\pt{\tilde{\delta}_\phi|N}\rightarrow\frac{P\pt{\tilde{\delta}_\phi|N}}{1-P_{\mathrm{fail}}}.
\end{equation} 
The cumulative distribution for the parameter $\tilde{\delta}_\phi$ with $N$ fixed is given by 
\begin{equation}
C_N\pt{\tilde{\delta}_{\phi,\mathrm{tr}}}=P\pt{\tilde{\delta}_\phi\leq\tilde{\delta}_{\phi,\mathrm{tr}}\vert N}=\sum_{\tilde{\delta_\phi=0}}^{\tilde{\delta}_{\phi,\mathrm{tr}}}P\pt{\tilde{\delta}_\phi|N}\label{CumDefinDe}
\end{equation}
and defines the probability of having a value of $\tilde{\delta}_\phi$ inferior to a threshold $\tilde{\delta}_{\phi,\mathrm{tr}}$.

\end{document}